\documentstyle[12pt,epsf]{article}

\newcommand{\gggg}{\gamma \gamma \to \gamma \gamma}

\begin{document}
\begin{flushright}
SLAC--PUB--8121\\
April 1999
\end{flushright}
\bigskip\bigskip

\def\grtsim{\,\,\rlap{\raise 3pt\hbox{$>$}}{\lower 3pt\hbox{$\sim$}}\,\,}
\def\lsim{\,\,\rlap{\raise 3pt\hbox{$<$}}{\lower 3pt\hbox{$\sim$}}\,\,}

\thispagestyle{empty}
\flushbottom

\centerline{\Large\bf$\gamma \gamma \to \gamma \gamma$ as a Test of Weak Scale Quantum Gravity at the NLC\footnote
{\baselineskip=14pt
Work supported by the Department of Energy, contract DE--AC03--76SF00515.}}
\vspace{22pt}

\centerline{\bf Hooman Davoudiasl}
\vspace{8pt}
  \centerline{\it Stanford Linear Accelerator Center}
  \centerline{\it Stanford University, Stanford, California 94309}
  \centerline{e-mail: hooman@slac.stanford.edu}
\vspace*{0.9cm}

\begin{abstract}

Recently, it has been proposed that the fundamental scale of quantum gravity can be close to the weak
scale if there are large extra dimensions .  This proposal has important phenomenological implications for
processes at the TeV scale.  We study the process $\gamma \gamma \to \gamma \gamma$, assuming an ultraviolet
cutoff $M_S \sim 1$ TeV for the effective gravity theory.  We find that, at center of mass energies $\sqrt s \sim 1$
TeV, the contribution of gravitationally mediated scattering to the cross section is comparable to that coming from
the one-loop Feynman diagrams of the Standard Model.  We thus conclude that the effects of weak scale quantum
gravity can be studied at the Next Linear Collider (NLC), in the  photon collider mode.  Our results suggest that, for
typical proposed NLC energies and luminosities, the range 1 TeV$\lsim M_S \lsim$ 10 TeV can be probed.

\end{abstract}

\newpage

\section{Introduction} 

The idea of using extra dimensions in describing physical phenomena is a fairly mature one and dates back to the
early decades of the twentieth century.  During that time, attempts at unifying the theories of electromagnetism and
gravitation were made by assuming the existence of an extra spatial dimension \cite{NKK}.  A new application of
extra dimensional theories has recently been proposed in Refs.\cite{ADD1, ADD2}, where it was suggested that the
fundamental scale of gravity $M_F$ could be as low as the weak scale $\Lambda_w \sim 1$ TeV, assuming that
there were $n$ large compactified extra dimensions.  It was shown in Refs.\cite{ADD1, ADD2} that gravitational data allow $n
\geq 2$.  This proposal has significant phenomenological implications for collider experiments at the scale
$\Lambda_w$, where Weak Scale Quantum Gravity (WSQG) effects are assumed to become strong.  Lately, a great deal of effort has
been made to constrain the proposal for WSQG \cite{Recent, SNGAM, KC}.  In the case of $n = 2$, the most stringent constraints
come from astrophysical and cosmological observations\cite{ADD2}, and it is argued that $M_F \grtsim 100$ TeV \cite{SNGAM}.
However, terrestrial experimental data have constrained WSQG to have $M_F \grtsim 1$ TeV, and in the case of $n
\geq 3$, there is no evidence of a more severe constraint.

In this paper, we show that the process $\gggg$ at energies of order 1 TeV can be used to constrain WSQG over a large range
in the TeV region.  This process can be studied at a proposed Next Linear Collider (NLC) \cite{NLC}, where high
energy Compton backward scattered photon beams with energies of order 1 TeV and luminosities of order 100
fb$^{-1}$ per year can be produced.  The process $\gggg$ has the advantage that it receives contributions from the
Standard Model (SM) only at the loop level and, therefore, could in principle be sensitive to new physics at the tree
level.  We will show that this process provides a good test of WSQG in the TeV regime.  

The rest of this paper will be organized as follows.  In section 2, we present the basic ideas of WSQG in theories
with large extra dimensions.  An estimate of the expected size of the gravity contribution to the photon
scattering process at weak scale energies will also be given in this section.  We conclude that the effect is strong, 
but the SM contribution could be
comparable and must be included.  Section 3 contains the SM and the gravity amplitudes used in our calculations. 
This section also includes our discussion of the approximations that have been made in writing down the various
amplitudes, and the conditions of their validity.  In section 4, we discuss the method used in computing the
predicted cross sections for $\gggg$ at the NLC, in the photon collider mode.  The results of our computations for
cross sections and the NLC reach for the effective scale of WSQG are presented in section 5.  Finally, section 6
contains our concluding remarks.

\section{WSQG and Its Contribution to $\gggg$}

In this work, we assume the fundamental scale of gravity $M_F \grtsim 1$ TeV and that there are $n \geq 2$ compact
extra dimensions of size $R$, even though there are astrophysical and cosmological considerations that suggest
$M_F \grtsim 100$ TeV for $n = 2$\cite{SNGAM}.  With these assumptions, Gauss' law yields the relation\cite{ADD1, ADD2}
\begin{equation}
M_P^2 \sim M_F^{n + 2} R^n, 
\label{MP} 
\end{equation} 
where $M_P \sim 10^{19}$ GeV is the Planck mass.  The exact relation among $M_P$, $M_F$, and $R$, as presented
in the appendix, depends on the convention and the compactification manifold used.  However, for order of
magnitude estimates, relation (\ref{MP}) suffices.

Given the above assumptions, we expect gravitationally mediated processes at TeV energies to be important.  To
estimate the size of the WSQG effect in $\gggg$, we take $M_F \sim 1$ TeV, and the center of mass energy
$\sqrt{s_{\gamma \gamma}} \sim 1$ TeV.  The gravity contribution $\sigma_G$ to the total cross section is then
given by
\begin{equation}
\sigma_G \sim \frac{2 \pi}{(16 \pi)^2} \left(\frac{E_\gamma}{M_F}\right)^6 \left(\frac{1}{M_F}\right)^2,
\label{sigG} 
\end{equation} 
and we obtain
\begin{equation}
\sigma_G \sim 10 \, {\rm fb},
\label{sigGfb} 
\end{equation}
where $E_\gamma = \sqrt{s_{\gamma \gamma}}/2$.  We note that, in the TeV regime, the SM total cross section
$\sigma_{SM} \sim 10$ fb is measurable at the NLC \cite{NLC, Jikia}, and Eq. (\ref{sigGfb}) suggests that the signal for
WSQG in $\gggg$ can be large and measurable, as well.  Our estimate also suggests that, although the effects of
gravity can be large, the SM contribution is comparable and has to be included in our analysis.  In the next section,
we present the SM and gravity amplitudes used in our calculations.

\section{The Amplitudes}

We consider the process $\gamma (k_1) \gamma (k_2) \to \gamma (p_1) \gamma (p_2)$, where $k_1$ and $k_2$ are the initial and
$p_1$ and $p_2$ are the final 4-momenta of the photons.  We define $s \equiv (k_1 + k_2)^2, t \equiv (k_1 - p_1)^2$, and $u \equiv
(k_1 - p_2)^2$.  Each photon can have either $+$ or $-$ helicity.  In what follows, we denote a helicity amplitude by $M_{i j k l}$, where
$i, j, k, l = \pm$, and $(i, j)$ are the helicities of the $(k_1, k_2)$ photons, and $(k, l)$ are the helicities of the $(p_1, p_2)$ photons. 
The 1-loop helicity amplitudes of the SM are in general complicated.  However, in the limit $s, |t|, |u| \gg m^2$, where $m$ is the
mass of a $W$ boson, a quark, or a charged lepton, these amplitudes can be approximated by those parts of them that receive
logarithmic enhancements\cite{Gounaris}.  Except for the contribution of the top quark loop which does not affect our results
significantly\cite{Gounaris}, these leading amplitudes provide a good approximation at the energies of interest to us, namely those
of the NLC.  We will discuss the necessary cuts and the regime of validity of these amplitudes, in the next section.

The process $\gggg$ has many symmetries that reduce the number of independent helicity amplitudes.  It can be shown
\cite{Gounaris} that only three helicity amplitudes, out of 16, are independent, and they are $M_{++++}(s, t, u), M_{+++-}(s, t, u)$, and
$M_{++--}(s, t, u)$.  In the high energy limit that we are considering, of these three amplitudes, the only logarithmically enhanced
one is $M_{++++}(s, t, u)$, for both the fermion and the $W$ loops\cite{Gounaris}.  For the $W$ loop amplitude we
have\cite{Gounaris}
\[ 
\frac{M_{++++}^{(W)}(s, t, u)} {\alpha^2} \approx 12 + 12 \left(\frac{u - t}{s}\right) \left[\ln \left(\frac{-u - i\varepsilon}{m_W^2}
\right) - \ln \left(\frac{-t - i\varepsilon}{m_W^2}\right) \right]
\]
\[
+ 16 \left(1 - \frac{3 t u}{4 s^2}\right)\left(\left[\ln \left(\frac{-u - i\varepsilon}{m_W^2}\right) - \ln \left(\frac{-t - i\varepsilon}
{m_W^2}\right)\right]^2 + \pi^2 \right)
\]
\[
+16 s^2 \left[\frac{1}{s t} \ln \left(\frac{-s - i\varepsilon}{m_W^2}\right) \ln \left(\frac{-t - i\varepsilon}{m_W^2}\right) +  
\frac{1}{s u} \ln \left(\frac{-s - i\varepsilon}{m_W^2}\right) \ln \left(\frac{-u - i\varepsilon}{m_W^2}\right) \right.
\]
\begin{equation}
+ \left.\frac{1}{t u} \ln \left(\frac{-t - i\varepsilon}{m_W^2}\right) \ln \left(\frac{-u - i\varepsilon}{m_W^2}\right)  \right],
\label{Wamp}
\end{equation}
where $\alpha \approx 1/137$ and $m_W$ is the mass of the $W$ boson; $m_W = 80$ GeV.  This value of $\alpha$ 
corresponds to that appropriate for real initial and final state photons\footnote{We thank I. Ginzburg for bringing this point to our attention.}. 

\newpage

For the fermion loops, we have 
\[ 
\frac{M_{++++}^{(f)}(s, t, u)}{\alpha^2 Q_f^4} \approx -8 - 8 \left(\frac{u - t}{s}\right) 
\left[\ln \left(\frac{-u - i\varepsilon}{m_f^2}
\right) - \ln \left(\frac{-t - i\varepsilon}{m_f^2}\right) \right]
\]
\begin{equation}
 - 4 \left(\frac{t^2 + u^2}{s^2}\right)\left(\left[\ln \left(\frac{-u - i\varepsilon}{m_f^2}\right) - \ln \left(\frac{-t - i\varepsilon}
{m_f^2}\right)\right]^2 + \pi^2 \right),
\label{famp}
\end{equation}
where $Q_f$ is the fermion charge in units of the positron charge, and $m_f$ is the mass of the fermion in the loop.  In our
approximation, there are only two more leading helicity amplitudes that will enter our computations.  These are 
\begin{equation}
M_{+-+-}(s, t, u) = M_{++++}(u, t, s)
\label{M+-+-}
\end{equation}
and  
\begin{equation}
M_{+--+}(s, t, u) = M_{+-+-}(s, u, t).
\label{M+--+}
\end{equation}

The gravity amplitudes are all at the tree level, and they are
\[
M^{(G, s)} = (2 \pi) \, \varepsilon^\rho(k_1) \, \varepsilon^\sigma(k_2) \,  \varepsilon^{* \gamma}(p_1) \, 
\varepsilon^{* \delta}(p_2) \, B^{\mu \nu, \alpha \beta}(k_1 + k_2) \, D(s)
\]
\begin{equation}
\times
[(k_1 \cdot k_2) C_{\mu \nu, \rho \sigma} + D_{\mu \nu, \rho \sigma}(k_1, k_2)] \, 
[(p_1 \cdot p_2) C_{\alpha \beta, \gamma \delta} + D_{\alpha \beta, \gamma \delta}(p_1, p_2)], 
\label{MGs}
\end{equation}
\[
M^{(G, t)} = (2 \pi) \, \varepsilon^\rho(k_1) \, \varepsilon^\sigma(k_2) \,  \varepsilon^{* \gamma}(p_1) \, 
\varepsilon^{* \delta}(p_2) \, B^{\mu \nu, \alpha \beta}(k_1 - p_1) \, D(t)
\]
\begin{equation}
\times
[(k_1 \cdot p_1) C_{\mu \nu, \rho \gamma} + D_{\mu \nu, \rho \gamma}(k_1, p_1)] \, 
[(k_2 \cdot p_2) C_{\alpha \beta, \sigma \delta} + D_{\alpha \beta, \sigma \delta}(k_2, p_2)], 
\label{MGt}
\end{equation}
and
\[
M^{(G, u)} = (2 \pi) \, \varepsilon^\rho(k_1) \, \varepsilon^\sigma(k_2) \,  \varepsilon^{* \gamma}(p_1) \, 
\varepsilon^{* \delta}(p_2) \, B^{\mu \nu, \alpha \beta}(k_1 - p_2) \, D(u)
\]
\begin{equation}
\times
[(k_1 \cdot p_2) C_{\mu \nu, \rho \delta} + D_{\mu \nu, \rho \delta}(k_1, p_2)] \, 
[(k_2 \cdot p_1) C_{\alpha \beta, \sigma \gamma} + D_{\alpha \beta, \sigma \gamma}(k_2, p_1)],
\label{MGu}
\end{equation}
where $\varepsilon^\mu (p)$ denotes the polarization vector of a photon with 4-momentum $p$, and the function $D(x)$ is 
given by\cite{Han}
\[
D(x) \approx M_S^{-4} \ln \left(\frac{M_S^2}{|x|}\right) \, \, \, \,  {\rm for} \, \, \, \, n = 2 
\]
and 
\begin{equation}
D(x) \approx M_S^{-4} \left(\frac{2}{n - 2}\right) \, \, \, \, {\rm for} \, \, \, \, n > 2;
\label{D(x)}
\end{equation}
the expressions for $B_{\mu \nu, \lambda \sigma}(k)$, $C_{\mu \nu, \lambda \sigma}$, and $D_{\mu \nu, \lambda \sigma}(k, p)$ are
given in the appendix.

Here, we would like to make a few comments regarding the amplitudes (\ref{MGs}), (\ref{MGt}), and (\ref{MGu}).  First, we note that the
expressions for $D(x)$ depend on the cutoff scale $M_S \gg s, |t|, |u|$, introduced to regulate the divergent sum over the infinite tower
of Kaluza-Klein states.  This dependence is a result of our implicit assumption that $M_S = M_F$.  However, if $M_S$ is much
smaller than $M_F$ then
\begin{equation}
D(x) \to \left(\frac{M_S}{M_F}\right)^{(n + 2)} D(x) \, \, \, \, {\rm for} \, \, \, \, n \geq 2,
\label{DMF}
\end{equation} 
resulting in a suppression\cite{Giudice}. 

Secondly, it should be kept in mind that the amplitudes (\ref{MGs}), (\ref{MGt}), and (\ref{MGu}) are derived from an effective
Lagrangian\cite{Han} with the lowest dimension operators that describe the coupling of the Kaluza-Klein gravitons to various fields,
in our case, the photon field.  In this effective description of quantum gravity, we need to introduce a cutoff $M_S \lsim M_F$, in
order to get finite results.  As with any effective Lagrangian, there are terms of higher dimension that should in principle be included in
the Lagrangian.  The term $\lambda (F_{\mu \nu} F^{\mu \nu})^2/M_S^4$, where $\lambda$ is an unknown coefficient, is one such term
that contributes at the same order in powers of $1/M_S$ to our calculations.  The coefficient $\lambda$ cannot be calculated, unless the
fundamental theory of gravity at scale $M_F$ is known.  In principle, the size of the contribution from this term can be larger than the
one calculated in this paper, and may even have the opposite sign.  However, since $\lambda$ is unknown, we have chosen to
consider only the lowest dimension local terms in the Lagrangian, and simply add the contributions from Eqs. (\ref{MGs}), (\ref{MGt}),
and (\ref{MGu}) to those obtained from Eqs. (\ref{Wamp}) and (\ref{famp}).  This is a reasonable choice, as long as one is only
interested in an order of magnitude estimate of the effects.  

\section{The NLC as a Photon Collider}

We mentioned before that high energy and luminosity $\gamma$ beams can be achieved at the NLC.  The basic proposed mechanism
uses backward Compton scattering of laser photons from the high energy $e^+ e^-$ beams at the NLC\cite{Ginzburg}.  The $\gamma$
beams that are obtained in this way have distributions in energy and helicity that are functions of the $\gamma$ energy and the initial
polarizations of the electron beams and the laser beams.  Laser beam polarization $P_l$ can be achieved close to $100\%$, however,
electron beam polarization $P_e$ is at the $90\%$ level.  We take $|P_l| = 1$ and $|P_e| = 0.9$ for our calculations.  

Let $E_e$ be the electron beam energy, and $E_\gamma$ be the scattered $\gamma$ energy in the laboratory frame.  The fraction of the 
beam energy taken away by the photon is then 
\begin{equation}
x = \frac{E_\gamma}{E_e}.
\label{x}
\end{equation}
We take the laser photons to have energy $E_l$.  Then, the maximum value of $x$ is given by 
\begin{equation}
x_{max} = \frac{z}{1 + z},
\label{xmax}
\end{equation}
where $z = 4 E_e E_l/m_e^2$, and $m_e$ is the electron mass.  One cannot increase $x_{max}$ simply by increasing $E_l$, since this
makes the process less efficient because of $e^+ e^-$ pair production through the interactions of the laser photons and the
backward scattered $\gamma$ beam.  The optimal value for $z$ is given by
\begin{equation}
z_{_{OPT}} = 2 \left(1 + {\sqrt 2}\right).
\label{zOPT}
\end{equation}
The photon number density $f(x, P_e, P_l)$ and average helicity $\xi_2 (x, P_e, P_l)$ are functions of $x$, $P_e$, $P_l$, and $z$, however, 
we always set $z = z_{_{OPT}}$ in our calculations.  We give the expressions for these two functions in the appendix. 

Let $M_{i j k l}$ be a helicity amplitude for $\gggg$.  We define 
\begin{equation}
|M_{++}|^2 \equiv \sum_{k, l} |M_{++ k l}|^2
\label{M++2}
\end{equation}
and 
\begin{equation}
|M_{+-}|^2 \equiv \sum_{k, l} |M_{+- k l}|^2,
\label{M+-2}
\end{equation}
where the summation is over the final state helicities of the photons.  Then, for various choices of the pairs $(P_{e_1}, P_{l_1})$ and
$(P_{e_2}, P_{l_2})$ of the the two beams, the differential cross section $d \sigma/d \Omega$ is given by
\[
\frac{d \sigma}{d \Omega} = \frac{1}{128 \, \pi^2 \, s_{ee}} \int \int d x_1 d x_2 \left[\frac{f(x_1) \, f(x_2)}{x_1 \, x_2}\right]
\]
\begin{equation}
\times
\left[\left(\frac{1 + \xi_2(x_1) \, \xi_2(x_2)}{2}\right)|M_{++}|^2 + \left(\frac{1 - \xi_2(x_1) \, \xi_2(x_2)}{2}\right)|M_{+-}|^2\right],
\label{diffcs}
\end{equation}
where $x_1$ and $x_2$ are the energy fractions for the two beams, given by Eq. (\ref{x}), and $s_{ee} = 4 E_e^2$.  
Different choices of $(P_{e_1}, P_{l_1})$ and $(P_{e_2}, P_{l_2})$ in $(f(x_1), \xi_2(x_1))$ and $(f(x_2), \xi_2(x_2))$,
respectively, yield different polarization cross sections.

We note that the expressions for $|M_{++}|^2$ and $|M_{+-}|^2$ are actually functions of the $\gamma \gamma$
center of mass energy squared ${\hat s} = x_1 x_2 s$, and the center of mass scattering angle $\theta_{cm}$.  We also
have ${\hat t} = x_1 x_2 t$ and ${\hat u} = x_1 x_2 u$.  In the previous section, we introduced the logarithmically
enhanced SM amplitudes, valid when $s, |t|, |u| \gg m_W^2$.  However, we see that to have a good approximation,
we must demand ${\hat s}, |{\hat t}|, |{\hat u}| \gg m_W^2$.  To avoid restricting the phase space too much, and in
order to have a good approximation to the SM amplitudes, we will make the following cuts
\[
\theta_{cm} \in [\pi/6, 5 \pi/6],
\]
\[
x_1 \in [\sqrt{0.4}, x_{1 max}],
\] 
and 
\begin{equation}
x_2 \in [\sqrt{0.4}, x_{2 max}],
\label{cuts}
\end{equation}
where $x_{1 max}$ and $x_{2 max}$ are given by Eq. (\ref{xmax}); $x_{1 max} = x_{2 max}$.  These cuts ensure
that the integrations are always performed in a region where ${\hat s}, |{\hat t}|, |{\hat u}| > m_W^2$. 

\section{Results}

In this section, we present our numerical results for the expected size of the WSQG effects at TeV energies. 
However, here, we would like to make a few remarks regarding our calculations.  First of all, in obtaining our
results, we have assumed $M_S = M_F$.  The effects of departure from this assumption are given in Eq. 
(\ref{DMF}).  Secondly, the only dependence on the number of extra dimensions $n$ in our computations comes
from Eq. (\ref{D(x)}).  We only distinguish between the cases with $n = 2$ and $n > 2$.  In the case with $n =
2$, in the limit $M_S^2 \gg s$, the WSQG amplitude is enhanced logarithmically compared to the case with $n >
2$.  In our computations, for $n > 2$, we have $\ln (M_S^2/{\hat s}) > 2/(n - 2)$ over most of the parameter
space considered.  We choose $n = 6$ as a representative value for $n > 2$; other choices result in a rescaling
of the effective value of $M_S$.  
  
\begin{figure}[htbp] 
\centerline{\epsfxsize=10truecm \epsfbox{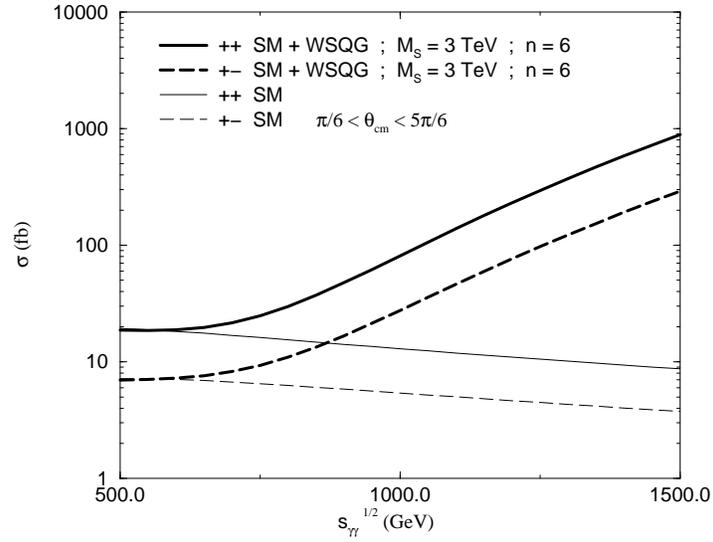}} 
\caption[1]{SM+WSQG and SM cross sections, represented by the thick and the thin lines, respectively, for
the initial helicities $++$ and $+-$.  The gravity contribution is calculated
for $M_S = 3$ TeV and $n = 6$.} 
\label{sgamma} 
\end{figure}

In Fig. (\ref{sgamma}), we present the $\gggg$ cross sections for SM + WSQG and SM, assuming a mono-energetic
beam of photons.  For the gravity contribution, we have chosen $M_S = 3$ TeV, and $n = 6$.  The curves in Fig.
(\ref{sgamma}) are obtained for photons with $++$ and $+-$ initial helicities.  At the NLC, the $\gamma$ beams
will have a distribution in photon energies and helicities, and these cross sections will not be observed. 
However, the cross sections presented in Fig. (\ref{sgamma}) show the relative size of the contribution of each
initial helicity state to the predicted cross section at the NLC, as obtained from Eq. (\ref{diffcs}).  

\begin{figure}[htbp]    
\centerline{\epsfxsize=10truecm \epsfbox{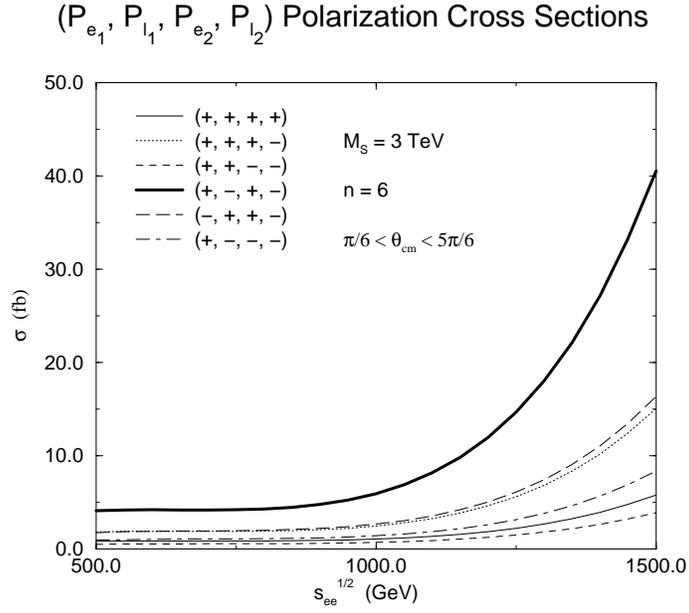}}
\caption[2]{SM + WSQG cross sections for six independent initial electron and laser beam polarizations.  Here, 
$M_S = 3$ TeV and $n = 6$.}
\label{sixpols}
\end{figure}

\begin{figure}[htbp]    
\centerline{\epsfxsize=10truecm \epsfbox{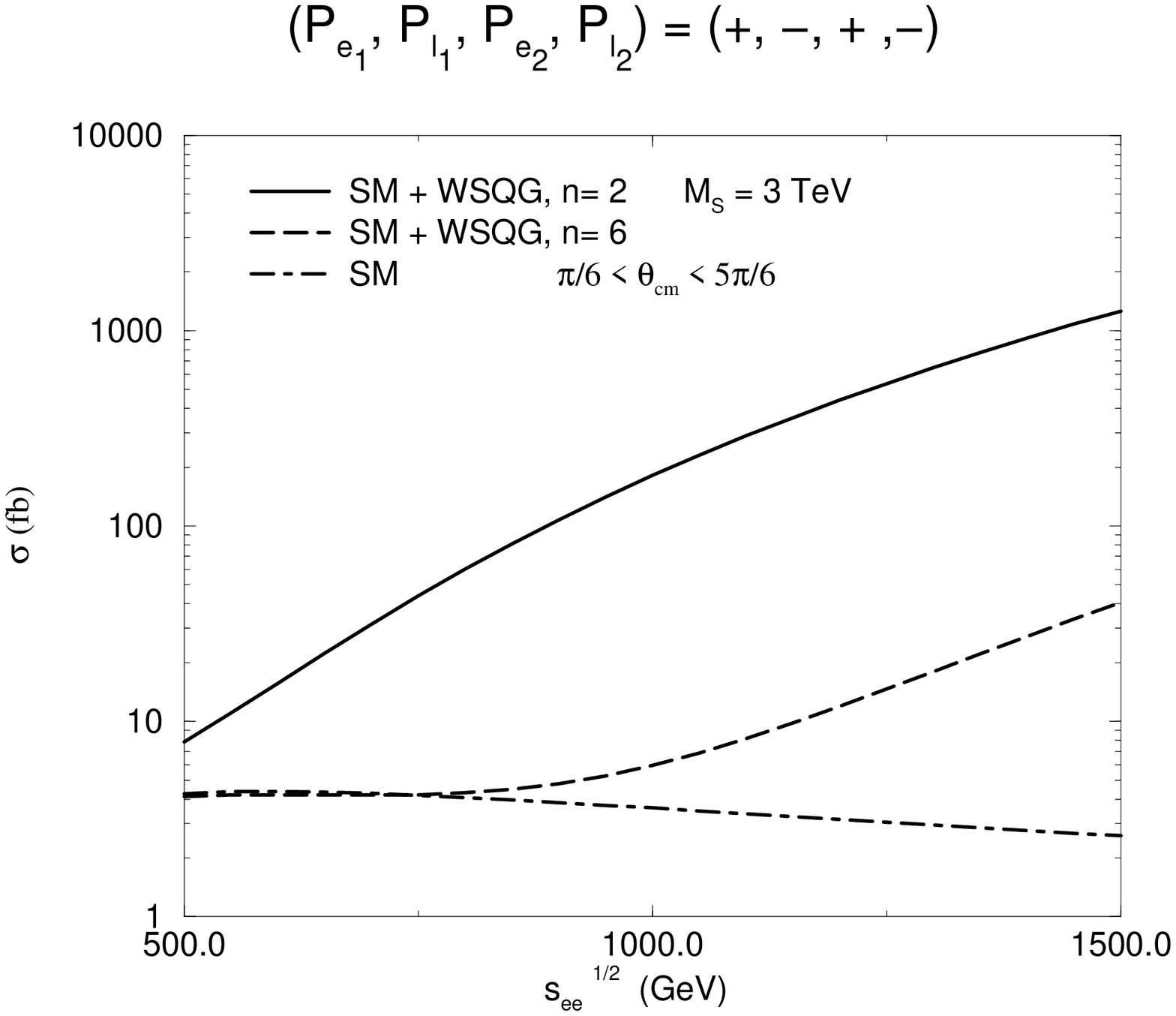}}
\caption[3]{SM + WSQG and SM cross sections for the $(+, -, +, -)$ polarization.  Here, 
$M_S = 3$ TeV and $n = 2, 6$, for the WSQG contributions.}
\label{26sm}
\end{figure}

The six SM + WSQG cross sections, for $M_S = 3$ TeV and $n = 6$, in Fig. (\ref{sixpols}), correspond to six independent choices
for the polarizations $(P_{e_1}, P_{l_1}, P_{e_2}, P_{l_2})$ of the electron and the laser beams at the NLC, in the photon collider
mode.  These cross sections are plotted versus the center of mass energy of the beam, $\sqrt{s_{ee}}$.  The curves in this figure
show a sensitive dependence on the choices of the polarizations for $\sqrt{s_{ee}} \grtsim 1$ TeV, with the $(+, -, +, -)$
polarization giving the largest cross section at high energies.  In Fig. (\ref{26sm}), choosing $M_S = 3$ TeV and $n =2,  6$, we
compare the SM + WSQG cross sections with that of the SM in the typical proposed NLC center of mass energy range
$\sqrt{s_{ee}} \in [500, 1500]$ GeV.  We have chosen the $(+, -, +, -)$ polarization for all three curves, since this choice yields the
largest gravity cross section, as shown in Fig. (\ref{sixpols}).  

\begin{figure}[htbp]    
\centerline{\epsfxsize=10truecm \epsfbox{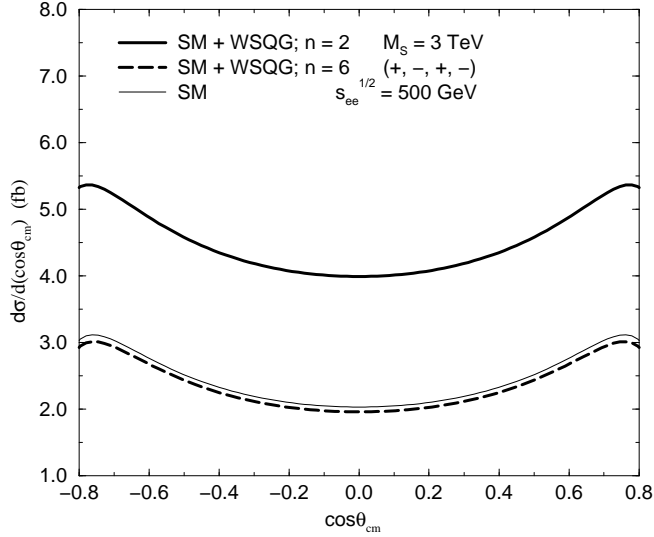}}
\caption[4]
{SM + WSQG and SM differential cross sections at $\sqrt{s_{ee}} = 500$ GeV for the $(+, -, +, -)$ polarization.  Here, $M_S = 3$ TeV and $n = 2,
6$, for the WSQG contributions.}
\label{dcs}
\end{figure}

The SM and SM + WSQG differential cross sections $d \sigma/d (\cos \theta_{cm})$, at $\sqrt{s_{ee}} = 500$ GeV, are compared in Fig.
(\ref{dcs}).  Again, we have chosen $M_S = 3$ TeV and the $(+, -, +, -)$ polarization.  The endpoint behavior of the differential cross sections
are caused by our choice for the cuts, given in Eqs. (\ref{cuts}).  At this center of mass energy, and given our cuts, the $n = 6$ result does not
offer a distinctive signal for WSQG.  For $n = 2$, we see that the differential cross section for SM + WSQG, in the region where $\cos
\theta_{cm} \approx 0$, is larger than that for SM by about a factor of 2.

Next, we present our results for the typical $M_S$ reach of the NLC, at various stages.  The stages at which $\sqrt{s_{ee}}
= 500$ GeV,  $\sqrt{s_{ee}} = 1000$ GeV, and $\sqrt{s_{ee}} = 1500$ GeV are denoted by NLC0.5, NLC1.0, and NLC1.5,
respectively.  Throughout, we assume that the luminosity $L = 100$ fb$^{-1}$ per year.  We use the $\chi^2 (M_S)$
variable, given by
\begin{equation}
\chi^2 (M_S) = \left(\frac{L}{\sigma_{_{SM}}}\right)\left[\sigma_{_{SM}} - \sigma (M_S)\right]^2,
\label{chi2}
\end{equation}
where $\sigma_{_{SM}}$ and $\sigma (M_S)$ refer to the cross sections  for the SM and SM + WSQG, respectively.  We have chosen the $(+, -,
+, -)$ polarization for computing $\sigma (M_S)$, since this choice gives the largest high energy SM + WSQG cross sections.  To establish the
reach in each case, we require a one-sided $95\%$ confidence level, corresponding to $\chi^2 (M_S) \geq 2.706$.

\begin{figure}[htbp]    
\centerline{\epsfxsize=10truecm \epsfbox{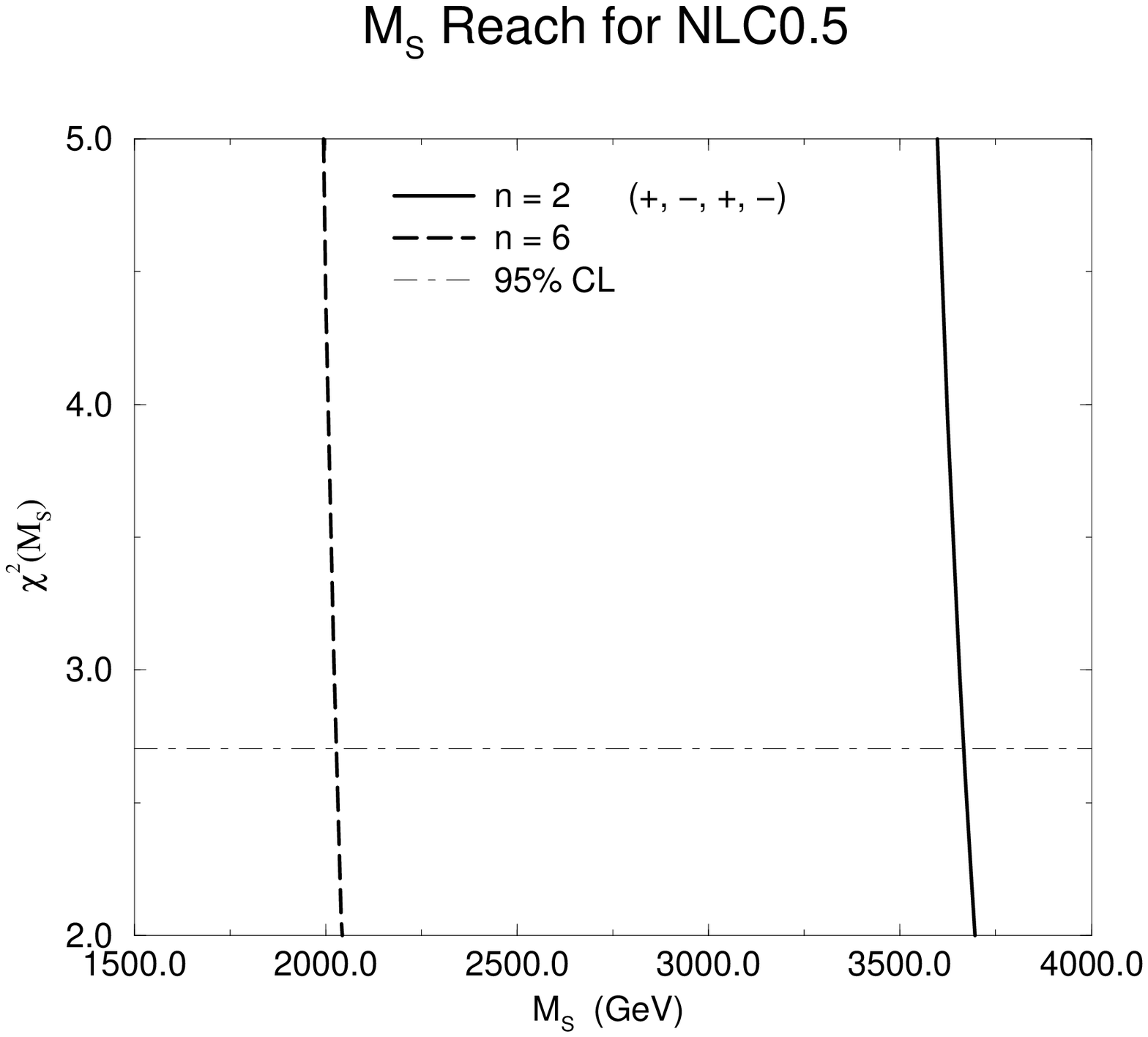}}
\caption[5]{The $M_S$ reach for NLC0.5.  The solid and the dashed lines represent the $\chi^2$ as a function of $M_S$ for the cases 
$n = 2$ and $n = 6$, respectively.  The dot-dashed line marks the reach at the $95\%$ confidence level.}
\label{reach0.5}
\end{figure}

\begin{figure}[htbp]    
\centerline{\epsfxsize=10truecm \epsfbox{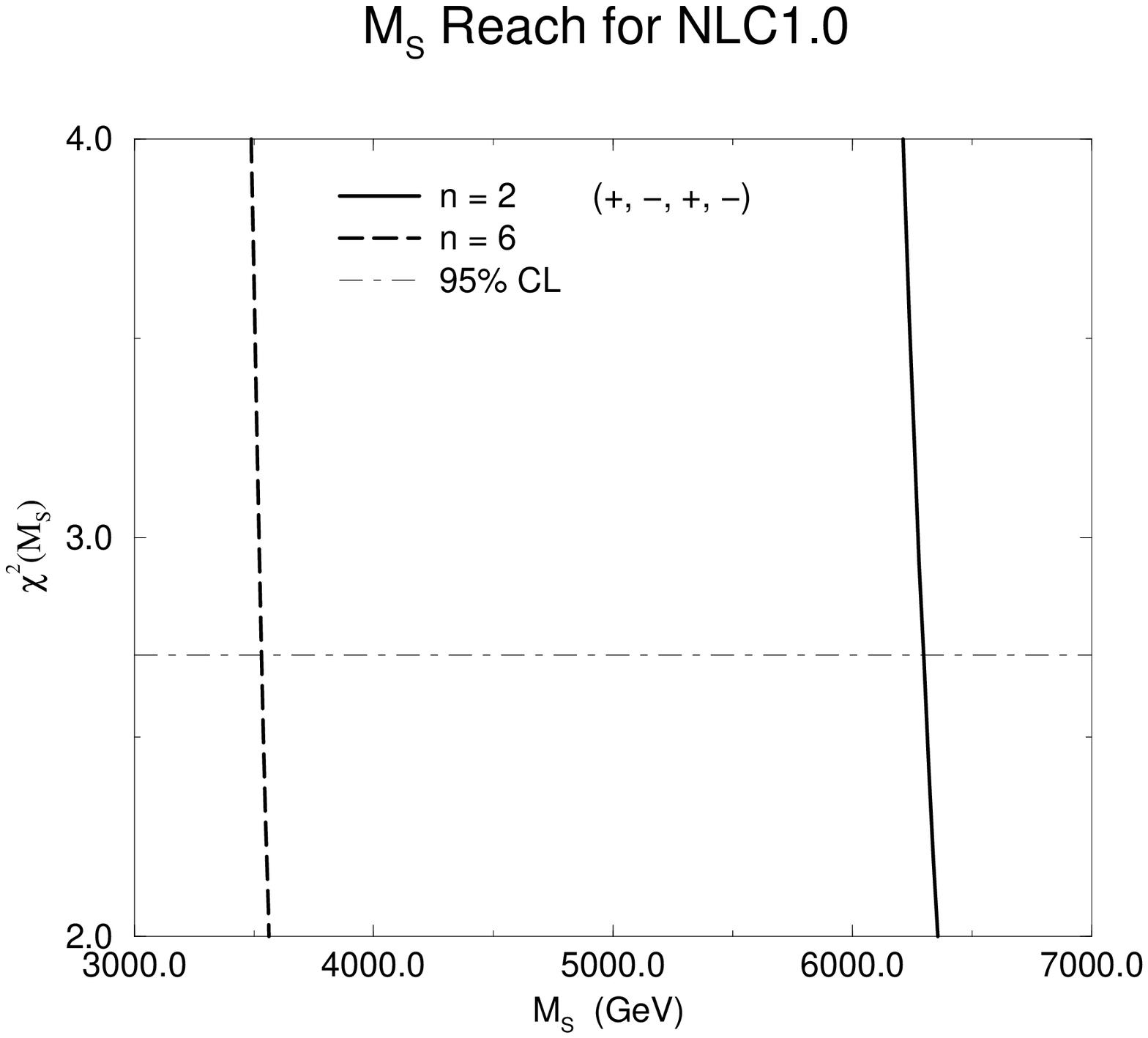}}
\caption[6]{The $M_S$ reach for NLC1.0.  The solid and the dashed lines represent the $\chi^2$ as a function of $M_S$ for the cases 
$n = 2$ and $n = 6$, respectively.  The dot-dashed line marks the reach at the $95\%$ confidence level.}
\label{reach1.0}
\end{figure}

\begin{figure}[htbp]    
\centerline{\epsfxsize=10truecm \epsfbox{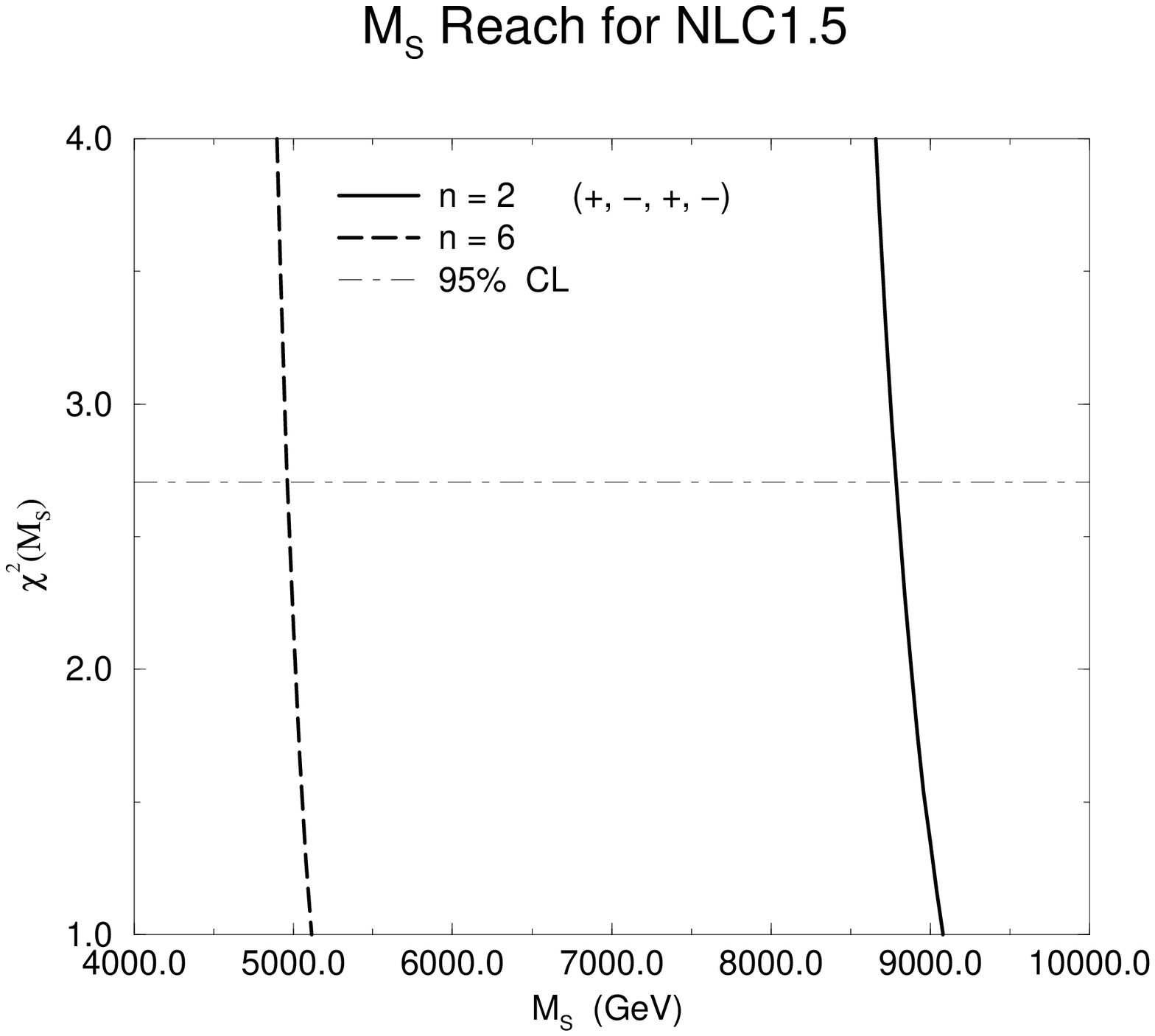}}
\caption[7]{The $M_S$ reach for NLC1.5.  The solid and the dashed lines represent the $\chi^2$ as a function of $M_S$ for the cases 
$n = 2$ and $n = 6$, respectively.  The dot-dashed line marks the reach at the $95\%$ confidence level.}
\label{reach1.5}
\end{figure}

The plots in Figs. (\ref{reach0.5}), (\ref{reach1.0}), and (\ref{reach1.5}) show the $95\%$ confidence level
experimental reach for $M_S$ at NLC0.5, NLC1.0, and NLC1.5, respectively.  As one can see from these figures,
the largest reach at each stage is for $n = 2$.  This is because of the logarithmic enhancement of the WSQG
amplitude for $n = 2$, as given by Eqs.  (\ref{D(x)}).  The lowest reach in $M_S$ is about 2 TeV for $n = 6$ at
NLC0.5 and the largest $M_S$ reach is about 9 TeV for $n = 2$ at NLC1.5.  Note that these values are obtained
for $L = 100$ fb$^{-1}$ per year, and by increasing $L$, the reach in $M_S$ will be improved.

\section{Concluding Remarks}

The process $\gggg$ at TeV energies is an important test channel for WSQG theories, since the tree level gravity contribution to
the process in these theories is expected to be significant, whereas the SM contributes only at the loop level.  In this paper, we
have used the high energy limit SM helicity amplitudes and the gravity amplitudes from the lowest dimension WSQG effective
Lagrangian to compute scattering cross sections.  The SM + WSQG cross sections can significantly differ from those of the SM
alone.

We have shown that the NLC in the photon collider mode can be effectively used to constrain theories of quantum gravity at the weak scale.  The size of the
expected effect shows a strong dependence on the choice of initial electron and laser polarizations.  Our computations suggest that studying $\gggg$ at the
NLC, operating at $\sqrt{s_{ee}} \in [500, 1500]$ GeV and $L = 100$ fb$^{-1}$ per year, can constrain the scale $M_S$ at which quantum gravity becomes
important, over the range 1 TeV $\lsim M_S \lsim$ 10 TeV.  

\section*{Acknowledgements}

It is a pleasure to thank T. Rizzo for many helpful discussions.  The author would also like to thank N. Arkani-Hamed, S. Brodsky, L. Dixon, I. Ginzburg,
J.L. Hewett, M. Peskin, J. Rathsman, and M. Schmaltz for various comments and conversations.  While this work was being completed, we received
a paper\cite{KC} by K.  Cheung whose contents have some overlap with those of this work.  

\appendix
\section*{Appendix}

In this paper, we have assumed that the fundamental mass scale $M_F$ of gravity and the size $R$ of the $n$ extra
dimensions are related by\cite{Han} 
\begin{equation}
\kappa^2 R^n = 16 \pi \, (4 \pi)^{n/2} \, \Gamma (n/2) \, M_F^{-(n + 2)},
\label{RMF}
\end{equation}
where $\kappa = \sqrt {16 \pi G_N}$; $G_N$ is the four dimensional Newton constant and $\Gamma$ represents the
Gamma-function.

The expressions for $B_{\mu \nu, \lambda \sigma}(k)$, $C_{\mu \nu, \lambda \sigma}$, and $D_{\mu \nu, \lambda \sigma}(k, p)$,
used in Eqs. (\ref{MGs}), (\ref{MGt}), and (\ref{MGu}), are given by\cite{Han}
\[
B_{\mu \nu, \lambda \sigma}(k) = \left(\eta_{\mu \lambda} - \frac{k_\mu k_\lambda}{m_{_{KK}}^2}\right) 
\left(\eta_{\nu \sigma} - \frac{k_\nu k_\sigma}{m_{_{KK}}^2}\right) + 
\left(\eta_{\mu \sigma} - \frac{k_\mu k_\sigma}{m_{_{KK}}^2}\right)
\left(\eta_{\nu \lambda} - \frac{k_\nu k_\lambda}{m_{_{KK}}^2}\right)
\]
\begin{equation}
-\frac{2}{3}\left(\eta_{\mu \nu} - \frac{k_\mu k_\nu}{m_{_{KK}}^2}\right)
\left(\eta_{\lambda \sigma} - \frac{k_\lambda k_\sigma}{m_{_{KK}}^2}\right),
\label{Bmnls}
\end{equation}

\begin{equation}
C_{\mu \nu, \lambda \sigma} = \eta_{\mu \lambda} \eta_{\nu \sigma} + \eta_{\mu \sigma} \eta_{\nu \lambda} - \eta_{\mu \nu} \eta_{\lambda \sigma},
\label{Cmnls}
\end{equation}
and
\begin{equation}
D_{\mu \nu, \lambda \sigma}(k, p) = \eta_{\mu \nu} k_\sigma p_\lambda - \left[\eta_{\mu \sigma} k_\nu p_\lambda + 
\eta_{\mu \lambda} k_\sigma p_\nu - \eta_{\lambda \sigma} k_\mu p_\nu + (\mu \leftrightarrow \nu)\right],
\label{Dmnls}
\end{equation}
respectively, where $\eta_{\mu \nu}$ is the Minkowski metric tensor and $m_{_{KK}}$ is the mass of a Kaluza-Klein state.  We have 
\begin{equation}
m_{KK}^2 = \frac{4 \, \pi^2 \, |{\vec n}|^2}{R^2},
\label{mKK}
\end{equation}
where ${\vec n} = (n_1, n_2, . . ., n_n)$, and $n_i$, $i = 1, 2, ..., n$, denotes the $n_i$ th Kaluza-Klein 
level in the $i$ th extra dimension. 

Let $P_e$ and $P_l$ be the polarizations of the electron beam and the laser beam, respectively.  We define the function $C(x)$\cite{Ginzburg} by
\begin{equation}
C(x) \equiv \frac{1}{1 - x} + (1 - x) - 4 r (1 - r) - P_e \, P_l \, r \, z (2 r - 1) (2 - x),
\label{C(x)}
\end{equation}
where $r \equiv x/[z(1 - x)]$.  Then, the photon number density $f(x, P_e, P_l; z)$ is given by
\newpage
\begin{equation}
f(x, P_e, P_l; z) = \left(\frac{2 \pi \alpha^2}{m_e^2 z \sigma_{_{C}}}\right) C(x), 
\label{f(x)}
\end{equation}
where 
\[
\sigma_{_{C}} = \left(\frac{2 \pi \alpha^2}{m_e^2 z}\right) \left[\left(1 - \frac{4}{z} -\frac{8}{z^2}\right) \ln (z + 1) + \frac{1}{2} + \frac{8}{z} - 
\frac{1}{2 (z + 1)^2}\right]
\]
\begin{equation}
+ P_e \, P_l \left(\frac{2 \pi \alpha^2}{m_e^2 z}\right) \left[\left(1 + \frac{2}{z}\right) \ln (z + 1) - \frac{5}{2} + \frac{1}{z + 1} - \frac{1}{2 (z + 1)^2}\right].   \label{sigC}
\end{equation}
The average helicity $\xi_2(x, P_e, P_l; z)$ is given by
\begin{equation}
\xi_2(x, P_e, P_l; z) = \frac{1}{C(x)}\left\{P_e \, \left[\frac{x}{1 - x} + x (2 r - 1)^2\right] - P_l \, (2r - 1)\left(1 - x + \frac{1}{1 - x}\right)\right\}.
\label{xi2}
\end{equation}

\end{document}